\documentstyle[sprocl]{article}

\def\Journal#1#2#3#4{{#1} {\bf #2}, #3 (#4)}

\def\PLB{{\em Phys. Lett.}  B}

\def\PRD{{\em Phys. Re
v.} D}

\def\mco{\multicolumn}

\def\ra{\rightarrow}

\def\ko{K^0}

\def\be{\begin{equation}}
\def\ee{\end{equation}}
\def\bea{\begin{eqnarray}}
\def\eea{\end{eqnarray}}

\begin{document}

\title{LEP, TOP and the STANDARD MODEL}

\author{ V.NOVIKOV,  \underline{L.OKUN} \footnote{Plenary talk at the
Second Sakharov Conference. Lebedev Physical Institute, Moscow, May
20-24, 1996.}, M.VYS
OTSKY}

\address{ITEP, Moscow, 117259, Russia}

\author{ A.ROZANOV}

\address{ITEP and CPPM, Marseille, France}

\maketitle\abstracts{
A simple way of deriving and analyzing electroweak radiative
corrections to the $Z$ boson decays is presented in the framework
of the Standard Model. The talk is based on a review article by the
authors "Electroweak radiative corrections in $Z$ decays" published
 in Uspekhi Fizicheskikh Nauk,  166 (1996) 539-574, in the May issue
 dedicated to 75th birthday o
f A.D.Sakharov. The  English translation
 of the article may be found on WWW (hep-ph 96 06 253).}

\section{LEP I and SLC}

LEP I (CERN) and SLC (Stanford) electron-positron colliders had
started their operation in the fall 1989 a few  months before
A.D.Sakharov passed away. The  sum of energies of $e^+ + e^-$ was
chosen to be equal to the $Z$ boson mass.  LEP I was terminated in the
fall of 1995  in order to give place to LEP II which alr
eady operates
at energy 135 GeV and will finally reach 192 GeV. SLC continues at
energy  close to 91 GeV.

The reactions, which has been  studied at LEP I (detectors: ALEPH,
DELPHI, L3, OPAL) and SLC(detector SLD) may be  presented in form:
$$
e^+e^- \to  Z \to f\bar{f}\;\;,
$$
where
\begin{center}

\vspace{3mm}

\begin{tabular}{ll}
$f\bar{f}$ = &
$\nu\bar{\nu}(\nu_e\bar{\nu}_e~,~~\nu_{\mu}\bar{\nu}_{\mu}~,
~~\nu_{\tau}\nu_{\tau})$ -- invisible~, \\
      & $l\bar{l}(e\bar{e}~, ~~\mu\bar{\
mu}~, ~~\tau\bar{\tau})$
      -- charged leptons~,\\
& $q\bar{q}(u\bar{u}~, ~~d\bar{d}~, ~~s\bar{s}~,~~ c\bar{c}~,
b\bar{b}) \to $  hadrons~.
\end{tabular}
\end{center}

\vspace{3mm}

About 20,000,000 $Z$ bosons
has been detected at LEP and about 100,000 at SLC (but here electrons
are polarized). Fantastic precision has been reached in the
measurement of the $Z$ boson properties:
$$
M_Z = 91,188.4 \pm 2.2
{\rm MeV}\;, \;\; \Gamma_Z = 2,496.3 \pm 3.2 {\rm MeV}\;,
$$
$$
\Gamma_h \equiv \Gamma_{hadrons} = 1,744.8 \pm 3.0 {\rm MeV}\;,
\;\; R_l = \Gamma_h/\Gamma_l = 20.788 \pm 0.032 \;,
$$
$$
\Gamma_{invisible} = 499.9 \pm 2.5 {\rm MeV}\;.
$$
By comparing the last number with  theoretical predictions for
neutrino decays it was established that the number of neutrinos which
interact with $Z$ boson is 3:
($N_{\nu} = 2.990 \pm 0.016$).
This is a result of fundamental importance.

More than 2.000 experimentalists and engineers and hundreds of
theorists participate in this unique collect
ive quest for truth!

\section{Theoretical analysis: the fundamental parameters.}

It is instructive to compare the electroweak theory with the quantum
electrodynamics (QED). In the latter there are two fundamental
parameters: mass of the electron, $m_e$, and its charge, $e$, or fine
structure constant $\alpha = e^2/4\pi$. Both  are not only
fundamental, but also known with high precision. Every observable in
QED can be expressed in terms of $m_e$ and $\alpha$ (and, of course,
of energies and mome
nta of particles participating in a given
process).

In the electroweak theory the situation is more complex for several
reasons:
\begin{enumerate}
\item{There are more fundamental charges and masses.}
\item{They are not independent of each other.}
\item{Not all of them are known with high accuracy.}
\item{Thus, as a basic parameter of the theory a
quantity is used, which is
known with highest accuracy, but which is not fundamental, the
four-fermion coupling of the muon decay, $G_{\mu}$.}
\end{enumerate}

The fundamental masses of the electroweak theory are masses of $W$ and
$Z$ bosons, $m_W$ and $m_Z$. Among the masses of fermions, the most
important for the $Z$-decay  is the mass of the top-quark, $m_t$.

The fundamental couplings of the electroweak theory are $e,~f,~g$, or
$\alpha = e^2/4\pi$, $\alpha_Z = f^2/4\pi$, $\alpha_W = g^2/4\pi$:
$e$ is the coupling of photons to electrically charged particles,
$f$ is the coupling of $Z$ bosons to weak neutral current, e.g.
$\bar{\nu}\nu$,

$g$ is the coupling of $W$ bosons to weak charged current, e.g.
$\bar{e}\nu$.

While the charged current is a purely V-A current of the form
$=\gamma_{\alpha}(1+\gamma_5)$, the ratio  $R_f$ between the vector
and axial vector neutral
currents depends on the third projection of the
isotopic spin of the fermion $f$, $T^f_3$, and on its electric
charge $Q^f$. The decay amplitude of the $Z$ boson decay may be
written in the form:
$$
M(Z \to f\bar{f}) = \frac{1}{2}f \bar{\psi}_f (g_{Vf} \gamma_{\alpha
}
+ g_{Af} \gamma_{\alpha}\gamma_5) \psi_f Z^{\alpha}\;\;,
$$
where $\psi_f$ -- is the wave function of emitted fermion, $\psi_f$
corresponds to the emitted antifermion (or absorbed fermion),
$Z^{\alpha}$ is the wave function of the $Z$ boson. At the tree level
$$
g_{Af} = T_3^f\;\;, \;\; g_{Vf} = T_3^f - 2Q^f s^2_W\;.
$$
Here
$$
T_3^f = +1/2 \;\;{\rm for} \;\; f = \nu,~ u,~ c\;; \;\;
T^f_3 = -1/2 \;\; {\rm for} \;\; f = l,~ d,~, s,~ b~.
$$
Thus
$$
R_f = g_{Vf}/g_{Af} = 1 - 4|Q^f| s^2_W\;\;.
$$
In the above expressions $s_W \equiv \sin \theta_W$, where $\theta_W$
-- is the so called weak angle. At the tree level (no loops):
$e/g = s_W\;\;, \;\;g/f = c_W\;\;, \;\; m_W/m_Z = c_W\;\;,
\;\;$ where $c_W \equiv \cos \theta_W$.

The four-fermion coupling constant $G_{\mu}$ is extracted from the
life-time of the muon, $\tau_{\mu}$, after taking into account
well-known electromagnetic corrections:
$$
1/\tau_{\mu} = \Gamma_{\mu} = \frac{G^2_{\mu} m^5_{\mu}}{192 \pi^3}
(1 + {\rm well known~~ 
corrections}\;
\sim \frac{m_e}{m_{\mu}}\;,\;
\alpha)\;\;.
$$
$$
G_{\mu} = (1.16639 \pm 0.00002) \cdot 10^{-5} {\rm GeV}^{-2}\;.
$$

In the tree approximation the four-fermion coupling constant
$G_{\mu}$ can be expressed in terms of $W$ boson coupling  constant
$g$ and its mass $m_W$:
$$
G_{\mu} = \frac{g^2}{4\sqrt{2}m^2_W} = \frac{\pi\alpha}{\sqrt{2}
m^2_W s^2_W} = \frac{\pi\alpha}{\sqrt{2} m^2_Z s^2_W c^2_W} \;.
$$
(The last two expressions are derived by using the relations $e/g =
s_W$, $\
alpha = e^2/4\pi$, $m_W/m_Z = c_W$).

\section{Theoretical analysis: the running $\alpha(q^2)$.}

It is well known since 1950's that  electric charge $e$ and hence
$\alpha$ logarithmically depend on the square of the four-momentum of
the photon, $q^2$. For a real photon $q^2 = 0$, for a virtual one
$q^2 \neq 0$. This phenomenon is usually referred to as "the running
of $\alpha$". It is caused by vacuum polarization, by loops of
virtual charged particles: charged leptons, $l\bar{l}$, and quarks,
$q
\bar{q}$, inserted into the propagator of a photon. As a result
$\bar{\alpha} \equiv \alpha(q^2 = m^2_Z)$ is approximately by 6\%
larger than $\alpha \equiv \alpha(0)$.

The relation between $\bar{\alpha}$ and $\alpha$ is obtained by
summing up an infinite series of insertions:
$\bar{\alpha} = \alpha/(1 - \delta\alpha)$;
$\delta\alpha = \delta\alpha_l + \delta\alpha_h$, where
$\delta\alpha_l$ is the one-loop contribution of leptons, while
$\delta\alpha_h$ -- is that of quarks (hadrons). The  lepton
ic
contribution can be predicted with very high accuracy. The hadronic
contribution is obtained on the basis of dispersion relations and
low-energy experimental data on $e^+e^-$-annihilation into hadrons.

The value of $\alpha(0)$ is known with very high accuracy:\\
$\alpha \equiv \alpha(0) = 1/137.035985(61)$;
$\alpha$ is very important for QED, but irrelevant to electroweak
physics.

The value of $\bar{\alpha}$ is less accurate:
$\bar{\alpha} = 1/128.896(90)$, but $\bar{\alpha}$ is
pivotal for
 electroweak physics. Let us stress
that the running of $\alpha$ is a purely electromagnetic effect,
caused by electromagnetic loops of light fermions. Contributions of
$t\bar{t}$ and $W^+W^-$ are negligibly small and may be taken into
account together with purely electroweak loops.

Unlike $\alpha(q^2)$, two other electroweak couplings $\alpha_Z(q^2)$
and $\alpha_W(q^2)$ are not running but "crawling" in the interval $0
\leq q^2 \leq m^2$:
$$
\alpha_Z(m^2_Z) = 1/22.91, \;\; \alpha_Z(0) = 1/23.10 
\;\;;
$$
$$
\alpha_W(m^2_Z) = 1/28.74, \;\; \alpha_W(0) = 1/29.01\;\;.
$$

The natural scale for $Z$-physics is $q^2 = m^2_Z$. Therefore it is
evident that $\bar{\alpha} \equiv \alpha(m^2_Z)$, not $\alpha \equiv
\alpha(0)$ is the relevant parameter. In fact, in all computer codes,
dealing with $Z$-physics, $\bar{\alpha}$  enters at a certain stage
and substitutes $\alpha$. But this occurs inside the "black box" of
the code, while $\alpha$ formally plays the role of an input
parameter.  In these 
codes the running of $\alpha$ is considered as
(the largest) electroweak correction. We consider this running as
purely electromagnetic one and define our Born approximation in terms
of $\bar{\alpha}$, $G_{\mu}$ and $m_Z$.

Instead of angle $\theta_W$, we define angle $\theta(s \equiv \sin
\theta\;, \;\; c \equiv \cos \theta$) by relation:
$$
G_{\mu} = \frac{g^2(q^2 = 0)}{4\sqrt{2} m^2_W} \simeq
\frac{e^2(m^2_Z)}{4\sqrt{2} s^2 m^2_W} =
\frac{\pi\bar{\alpha}}{\sqrt{2} s^2 c^2 m^2_Z}\;\;,
$$
where
 the second equality is based on the "crawling" of $g(q^2)$:
$g(0) \simeq g(m^2_Z)$.
Thus,
$$
\sin^2 2\theta = \frac{4\pi\bar{\alpha}}{\sqrt{2} G_{\mu} m^2_Z} =
0.71078(50)\;\;,
$$
$$
s^2 = 0.23110(23)\;\;,\;\; c^2 = 0.76890(23)\;\;,
\;\; c = 0.87687(13)\;\;.
$$
Our Born approximation starts with the most accurately known
observables: $ G_{\mu}\;, \; m_Z\;, \; \bar{\alpha}\;\; ({\rm or}\;
s^2)\;$.

The traditional parametrization of electroweak theory in terms of
$G_{\mu}\;, \; \alpha$, and 
$ s^2_W \equiv 1 - m^2_W/m^2_Z$ is less
convenient ($s_W$ has poor accuracy: $\Delta m_W = \pm 160$ MeV;
running of $\alpha$ is not separated from electroweak corrections and
overshadowes them.)

\section{Theoretical analysis: one-loop electroweak corrections.}

For the sake of brevity let us choose two observables:
$$
s^2_W \equiv 1 - \frac{m^2_W}{m^2_Z} \;\;, \;\;
s^2_l \equiv \frac{1}{4}(1 - \frac{g_{Vl}}{g_{Al}}) \equiv
\frac{1}{4}(1 - R_l)\;\;.
$$

In the Born approximation $s^2_W = s^2_l
 = s^2$. From UA2 and CDF
experiments:
$$
s^2_W = 0.2253(31)\;\;,
2\sigma \;\;{\rm away~~ from}\;\; s^2 = 0.23110(23)\;.
$$
From LEP and SLC:
$$
s^2_l = 0.23141(28)\;, \;\; 1\sigma \;\;{\rm away~~ from}\;\; s^2\;.
$$
(Note the high experimental accuracy of $s^2_l$ compared to that of
$s^2_W$.)

In the one-loop approximation
$$
s^2_l = s^2 - \frac{3}{16\pi} \frac{\bar{\alpha}}{c^2 - s^2} V_{R_l}
(m_t,\; m_H)\;\;,
$$
where $c^2 - s^2 = 0.5378$ and the radiative correction depends on
the mas
ses of the top quark and higgs.
These masses enter via loops containing virtual top quark, or higgs.
The coefficient in front of $V_{R_l}$ is chosen in such a way that
$V_{R_l}(m_t\;,\; m_H) \approx t \equiv (\frac{m_t}{m_Z})^2\;\;.
$
The same asymptotic normalization is used for the radiative
corrections to other electroweak observables.
The good agreement, within $1 \div 2\sigma$, between experimental
values of $s^2_W\;, \; s^2_l$ and their Born values means that
electroweak radiative corrections
 are anomalously small. The
unexpected smallness of $V_{R_l}$ is the result of cancellation
between large and positive contribution from the $t$-quark loops and
large and negative contribution from loops of other virtual
particles. This cancellation, which looks like a conspiracy, occurs
when $m_{top}$ is around 160 GeV, if higgs is light $(m_H \leq 100 $
GeV). If higgs is heavy $(m_H = 1000$ GeV) it occurs when $m_{top}$
is around 210 GeV. Thus, vanishing electroweak radiative corrections
tell us t
hat top is heavy.

\section{LEPTOP and the general fit.}

The analytical formulas for all electroweak observables have been
incorporated in our computer code which we dubbed LEPTOP.
The fit of all electroweak data by LEPTOP gives:
$$
m_t = 180 \pm 7^{+7}_{-21}\; {\rm GeV}\;.
$$
The central value $(180 \pm 7)$ corresponds to $m_H = 300$ GeV; the
shifts (+18, -21) -- to $m_H = 1000$ and 60 GeV, respectively.
This prediction is in perfect agreement with the recent (spring 1996)
data on the direct 
measurements of the top mass by two collaborations
at FNAL:
$$
m_t = 175.6 \pm 5.7 \pm 7.4 \; {\rm GeV \;\;\; (CDF)}\;\;, m_t =
     170 \pm 15 \pm 10 \; {\rm GeV \;\;\; (D0)}\;\;.
$$
(Here the first uncertainty is statistical, the second -- systematic
one.)

Electroweak radiative corrections depend on $\ln m_H/m_Z$ and give at
present unreliable limits on $m_H$.

Hadronic decays of $Z$ are sensitive to the value of the gluonic
coupling $\alpha_s$:

$$
\Gamma_q \equiv \Gamma(Z \ra q\bar{q}) 
= 12\Gamma_0 [g^2_{Aq}
R_{Aq} + g^2_{Vq} R_{Vq}]\;,
$$
where
$$
\Gamma_0 = \frac{G_{\mu} m^3_Z}{24\sqrt{2} \pi} = 82.944(6) \;{\rm
MeV}\;\;,
$$
and the "radiators" $R_{Aq}$ and $R_{Vq}$ contain QCD corrections
caused by emission and exchange of gluons. In the first approximation
$R_{Vq} = R_{Aq} = 1 + \frac{\hat{\alpha}_s}{\pi}\;\;,
$
where $\hat{\alpha}_s \equiv \alpha_s(m^2_Z)$ in the $\overline{MS}$
scheme. The LEPTOP fit of all electroweak data gives:
$ \hat{\alpha}_s = 0.124(4)^{+2}_{-2}\
;\;.$
Here the central value $(0.124\pm 0.004)$ corresponds to $m_H = 300$
GeV; the shifts $+0.002 $
and $-0.002$ -- to $m_H = 1000$ GeV and 60 GeV,
respectively.

Let us note, that low energy processes (deep inelastic scattering,
$\Upsilon$-spec\-tro\-sco\-py) give much smaller values of
$\hat{\alpha}_s$, around 0.110, when extrapolated to $q^2 = m^2_Z$.
There are different opinions on the seriousness of this discrepancy.

Another problem  is connected with the experimental value of the
width o
f the decay $Z \to b\bar{b}$. Theoretically the ratio $R_b =
\Gamma_b/\Gamma_h$ is not sensitive to $\hat{\alpha}_s$, $m_t$ and
$m_H$; the theory predicts:
$
R_b = 0.2155(3)^{-7}_{+7}\;\;,
$
where again the central value $(0.2155 \pm 0.0003)$ corresponds to
$m_H = 300$ GeV, and  shifted by -0.0007 for $m_H = 1000$ GeV and
by $+0.0007$ for $m_H = 60$ GeV. Experimentally $R_b = 0.2219(17)$,
which is more than $3\sigma$ larger than the theoretical prediction
based on the Standard Model.

Both probl
ems (of $\hat{\alpha}_s$ and of
$R_b$) may be solved by new
physics. Theorists try to change their predictions by considering:
 a) the existence of light supersymmetric particles (squarks, winos,
gluinos), which contribute to electroweak loops;
 b) the existence of another $Z$ boson -- $Z'$, which is more strongly
coupled to $b\bar{b}$, than to $e\bar{e}$ ("beautyphilic and
leptophobic".)
But maybe, experimentalists, can also change their numbers?

\end{document}